\documentclass[fleqn,10pt]{wlscirep}
\usepackage[utf8]{inputenc}
\usepackage[T1]{fontenc}
\title{Effects of Population Co-location Reduction on Cross-county Transmission Risk of COVID-19 in the United States}

\author[1,*]{Chao Fan}
\author[2]{Sanghyeon Lee}
\author[3]{Yang Yang}
\author[3]{Bora Oztekin}
\author[1]{Qingchun Li}
\author[1,*]{Ali Mostafavi}
\affil[1]{Zachry Department of Civil and Environmental Engineering, Texas A\&M University, College Station, TX 77843, U.S.}
\affil[2]{Department of Electrical and Computer Engineering, Texas A\&M University, College Station, TX 77843, U.S.}
\affil[3]{Department of Computer Science and Engineering, Texas A\&M University, College Station, TX 77843, U.S.}
\affil[*]{chfan@tamu.edu; amostafavi@civil.tamu.edu}

\begin{abstract}
The objective of this study is to examine the transmission risk of COVID-19 based on cross-county population co-location data from Facebook. The rapid spread of COVID-19 in the United States has imposed a major threat to public health, the real economy, and human well-being. With the absence of effective vaccines, the preventive actions of social distancing, travel reduction and stay-at-home orders are recognized as essential non-pharmacologic approaches to control the infection and spatial spread of COVID-19. Prior studies demonstrated that human movement and mobility drove the spatiotemporal distribution of COVID-19 in China. Little is known, however, about the patterns and effects of co-location reduction on cross-county transmission risk of COVID-19. This study utilizes Facebook co-location data for all counties in the United States from March to early May 2020 for conducting spatial network analysis where nodes represent counties and edge weights are associated with the co-location probability of populations of the counties. The analysis examines the synchronicity and time lag between travel reduction and pandemic growth trajectory to evaluate the efficacy of social distancing in ceasing the population co-location probabilities, and subsequently the growth in weekly new cases across counties. The results show that the mitigation effects of co-location reduction appear in the growth of weekly new confirmed cases with one week of delay. The analysis categorizes counties based on the number of confirmed COVID-19 cases and examines co-location patterns within and across groups. Significant segregation is found among different county groups. The results suggest that within-group co-location probabilities (e.g., co-location probabilities among counties with high numbers of cases) remain stable, and social distancing policies primarily resulted in reduced cross-group co-location probabilities (due to travel reduction from counties with large number of cases to counties with low numbers of cases). These findings could have important practical implications for local governments to inform their intervention measures for monitoring and reducing the spread of COVID-19, as well as for adoption in future pandemics. Public policy, economic forecasting, and epidemic modeling need to account for population co-location patterns in evaluating transmission risk of COVID-19 across counties.
\end{abstract}
\begin{document}

\flushbottom
\maketitle
%
%
\section*{Introduction}
The coronavirus disease 2019 (COVID-19) has caused a pandemic which threatens public health, the economy, and human well-being\cite{Mehta2020,Fan2020b}. As of 18 May 2020, more than 4.7 million people worldwide have been infected, with 1.5 million cases being confirmed in the United States\cite{Google}. In fact, United States has suffered the greatest number of confirmed cases in the world. Given the absence of effective vaccines and drugs, non-pharmacologic measures are essential to control the spread of COVID-19 in the United States\cite{Gao2020}.

Social distancing is one non-pharmacologic interventions adopted to reduce the transmission of COVID-19\cite{Caley2008}. In particular, state and local governments in the United States have issued stay-at-home orders, discouraged air travel, and closed non-essential businesses\cite{Mervosh2020}. Schools, from preschool through higher education have closed, with most classes resuming from electronic platforms. These measures were enacted in an effort to reduce person-to-person contact and its resulting close contact of people from different regions. Existing studies have demonstrated that cross-region travel drives the spatiotemporal distribution of COVID-19\cite{Jia2020,Oliver2020}. Modeling of travel restrictions in the most heavily affected regions was projected to be successful in slowing overall epidemic progression and reducing the transmission of the SARS-COV-2 virus in China\cite{Chinazzi2020}. In addition, recent studies have proposed and tested multiple mathematical disease-spread models, such as susceptible-infectious-recovery (SIR) model\cite{Liu2018a,Newman2002} and its derived models\cite{Giordano2020,Prem2020,Aleta2020,Ogbunugafor2020,Fan2020d}, and the global epidemic and mobility model (GLEAM)\cite{Balcan2009}, to evaluate the trajectories of virus spread and the effectiveness of intervention measures. Travel reduction, however, is not always well accounted for in the United States\cite{Gallotti2020}. Due to the varying awareness of transmission risks, people in different regions may respond to the COVID-19 pandemic in different manners. The variation of cross-county population co-location and travel reduction may lead to inadequate outcomes, deviate from model predictions, and even trigger the transmission of the disease in some regions. Hence, the fight against the spread of COVID-19 requires an empirical quantitative and grounded assessment of the cross-county population co-location patterns and effects of co-location reduction (stay-at-home order) on the transmission risks. 

A large-scale, systematic analysis of the population co-location and travel reduction in the United States is now feasible thanks to Facebook weekly co-location maps, which estimate the extent to which people from different regions are co-located for all counties. Data from the first week of March to the first week of May 2020 is available. Facebook co-location maps enable studying the cross-county transmission risks due to travel patterns among people from different regions, which is essential for modeling the transmission of diseases across regions. Through quantifying the mixing patterns of people from different regions, Facebook co-location maps offer an intuitive parametrization for calculating co-location probabilities with temporal fluctuations over the course of COVID-19 pandemic\cite{FacebookDataforGood2020} (See Materials and Methods for more details).

In this study, through the transformation of Facebook co-location maps to spatial networks, we examined the transmission risk patterns across different counties, how co-location probabilities are reduced proactively due to stay-at-home orders, and the effect of travel reduction on the spatiotemporal transmission of COVID-19 in the United States. In our spatial network model, nodes represent counties and the edge weights represent co-location probabilities. Accordingly, we characterized the cross-county transmission risks based on the co-location degree centrality of each county and determined the reduction in travel based on the reduction in the co-location degree centrality. Our analysis is based on the time series of weekly co-location reduction and number of weekly new confirmed cases, as well as the county-level basic reproduction numbers (obtained from estimating the parameters of the simple epidemic model based on the number of confirmed cases). We analyzed the synchronicity between the time series using dynamic time warping and quantified the time lags between the two metrics. Our results indicate that adherence to social distancing policy and a halt to all nonessential travel positively mitigate the growth rate of weekly new cases (second-order growth rate) and the estimated basic reproduction number, but the mitigation effects appeared with a one-week delay in some counties. To account for the variation in the transmission risks in different counties, we further grouped the counties into three categories based on the size of the infected population and studied the co-location changes and travel reduction patterns within and across groups. Highly infected (large number of infections) counties usually have large populations, which are hotspots with greater risk of contamination. We also found segregation in the co-location of populations among the counties in different groups, which indicates a reduction in co-location probabilities between counties with large numbers of cases and counties with a fewer number of cases. This segregation contributed to blocking the transmission of COVID-19 from highly infected counties to other counties. The co-location probabilities among counties with a large number of cases, however, remained stable, which could negatively affect the ability to contain the spread of the virus in highly infected counties.

\section*{Results}
\subsection*{Co-location degree centrality as a metric of cross-county travel} 
\begin{figure}[ht]
\centering
\includegraphics[width=17cm]{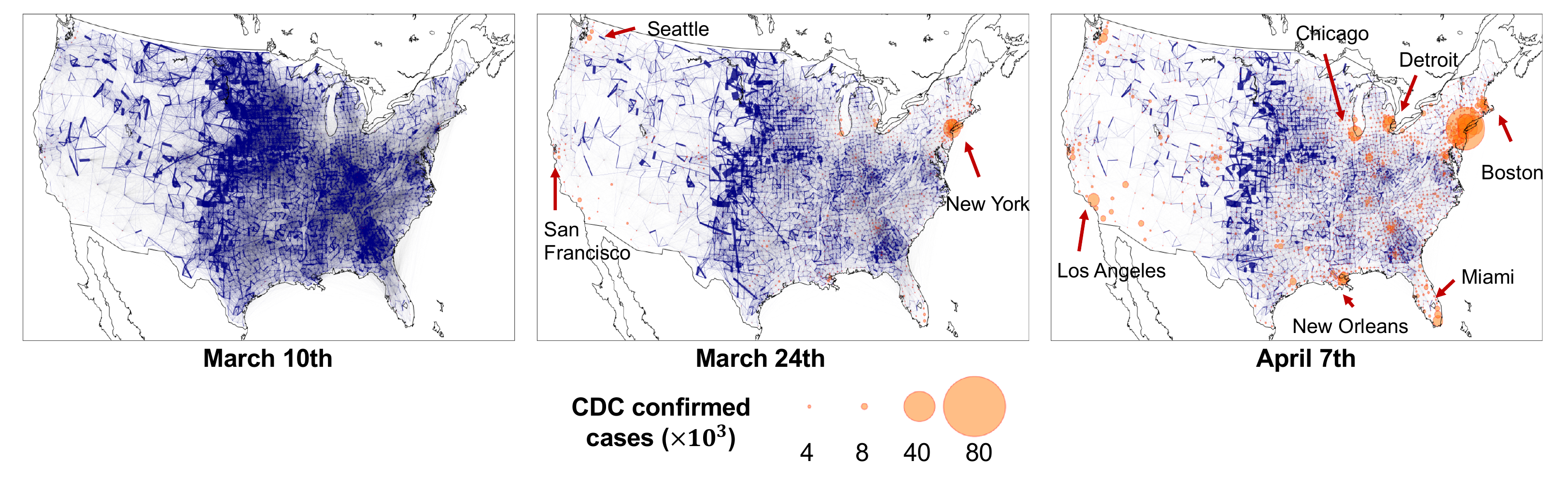}
\caption{Population co-location probability among US counties. A node represents a county; an edge represents the co-location relationship between two counties. The size of a node is proportional to the number of confirmed cases in a county; the width of an edge is proportional to the co-location probability of two counties.}
\label{fig_1}
\end{figure}

The first case of COVID-19 was confirmed in the United States in late January 2020; the pandemic broke out in March 2020. Fig. \ref{fig_1} shows the number of confirmed cases by county. Intuitively, we find that the majority of the confirmed cases are present in counties with large population sizes and population densities. It is also evident that the number of confirmed cases grew rapidly during March and April as awareness and testing capacity increased. With the increased awareness of the virus transmission risks, many state and local governments started to issue the stay-at-home order in mid-March\cite{Mervosh2020}. Accordingly, travel was drastically reduced during late March and April.

Facebook co-location data can inform about the patterns of reduced travel between people from different pairs of counties. Fig. \ref{fig_1} visualizes the co-location probability for each pair of counties in the U.S. Since calculation of co-location probability takes into account the population size of a county (see Materials and Methods for details), the co-location probability between two metropolitan areas is relatively low. The high co-location probabilities tend to appear on the edges between two contiguous counties in rural or sparsely populated areas, such as the counties in the Midwest. In addition, we can observe that co-location probabilities decreased over time as the COVID-19 outbreak in the continued to grow.

To characterize the travel connections of a county with other counties, we adopted the degree centrality concept from network science to aggregate the co-location probability for each county (See Materials and Methods)\cite{Fan2019}. Degree centrality indicates the extent to which people in a county have contact with people from other counties through cross-county travel. As the virus is usually transmitted by human contact, the higher the co-location degree centrality of a county, the higher the probability that the novel coronavirus would be transmitted to or from other counties. The reduction in degree centrality can reflect the reduction of travel across the counties. As a part of social distancing, travel reduction contributed to the mitigation of virus transmission. Hence, this study focuses on examining the fluctuations in degree centrality of counties to enable a quantitative assessment of social distancing and its relationship with the spatiotemporal transmission risk of COVID-19 in the U.S.

To quantify the severity of COVID-19 in U.S. counties, we used the number of confirmed cases for each county from the US Centers for Disease Control and Prevention (CDC) (see Materials and Methods). Since the Facebook co-location data was computed weekly, to be consistent, we also calculated the cumulative confirmed cases by week. The outcomes of disease transmission include not only the number of confirmed cases, but also new cases confirmed each week, which is usually characterized by the basic reproduction number, first-order growth rate, and second-order growth rate. The basic reproduction number ($R_0$) is defined as the expected number of secondary cases generated by a single infection in a completely susceptible population\cite{Dietz1993}. The basic production number is the epidemic threshold of the transmission of a pandemic and can be reduced through measures such as social distancing. We estimated $R_0$ based on a simple epidemic model and the growth of confirmed cases over time. (See Materials and Methods). The first-order growth rate in this study is defined as the number of weekly new cases in each county, which is affected by the number of confirmed cases which become the infection sources and by the population contacts which could expedite the transmission process. The first-order growth rate is further driven by the second-order growth rate which is defined by the changes of first-order growth rate (weekly new cases). Population travel would increase the co-location transmission probability, basic reproduction number, and further boost the number of weekly new cases. 

\begin{figure}[!htb]
\centering
\includegraphics[width=17cm]{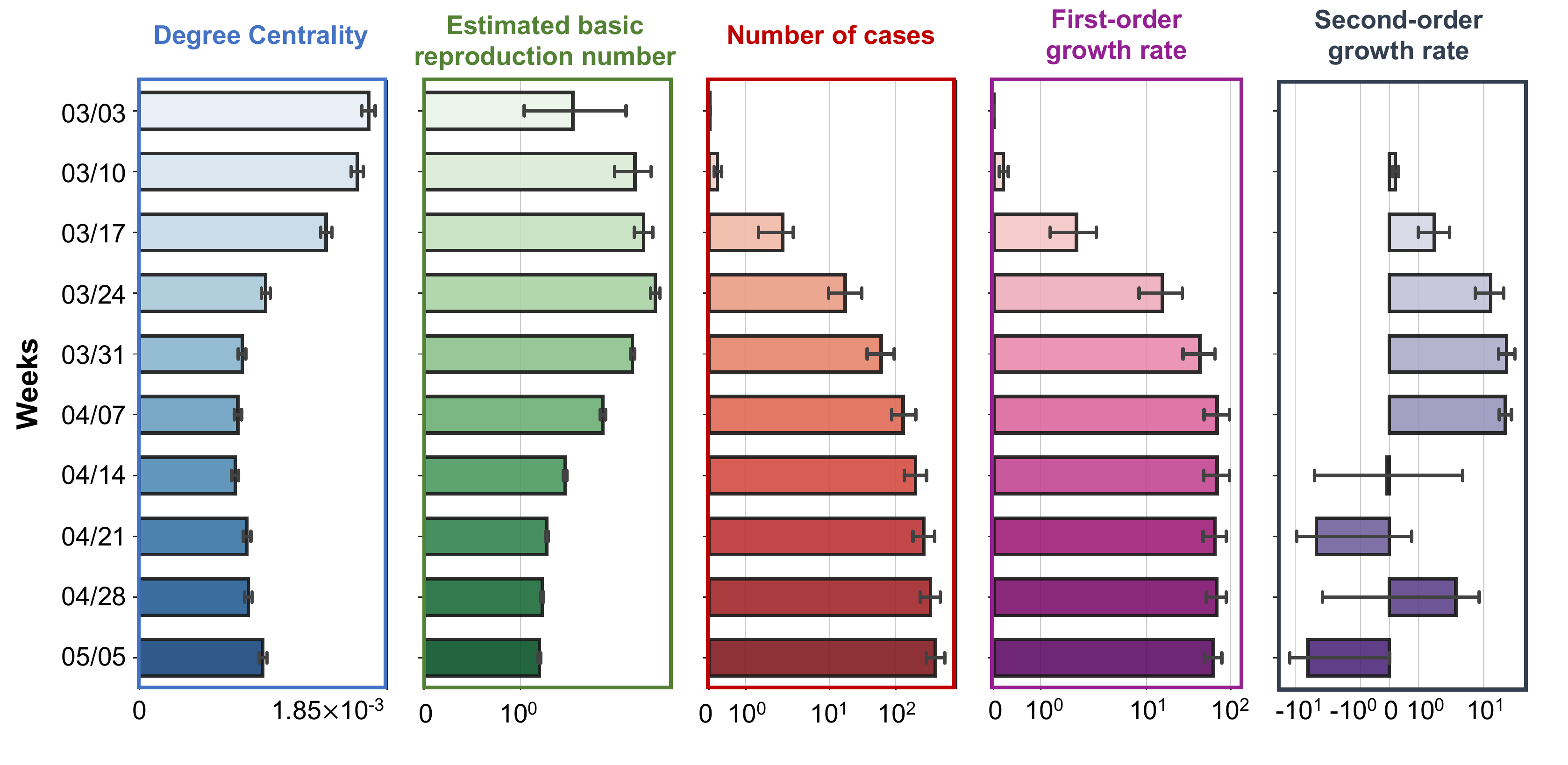}
\caption{Five metrics for measuring population co-location and COVID-19 situation over time: co-location degree centrality, estimated basic reproduction number, number of cases, first-order growth rate, and second-order growth rate.}
\label{fig_2}
\end{figure}

Fig. \ref{fig_2} shows the time series of all metrics adopted in this study. We can see that the co-location degree centrality decreased dramatically in March and remained low in April, while the number of cases increased in March and April. The weekly new confirmed cases (first-order growth rate) increased in March, peaked in early April, and decreased slightly in late April. The change in the weekly new cases is quantified by the second-order growth rate (fourth column in Fig. \ref{fig_2}). In March, the second-order growth rate increased rapidly but decreased and even dropped below zero in mid- and late April. The basic reproduction number grew along with the increase in cases at the early stages of the outbreak, but decreased following the reduction of the degree centrality. These observations raise two important questions regarding the temporal relationships between the reduction of co-location degree centrality: the second-order growth rate, and the basic reproduction number. Understanding these relationships could be beneficial for capturing the effect of population and travel reduction (stay-at-home orders) on the dynamics of the infection growth and for developing effective non-compulsory measures and epidemic models. 

\begin{figure}[!htb]
\centering
\includegraphics[width=17cm]{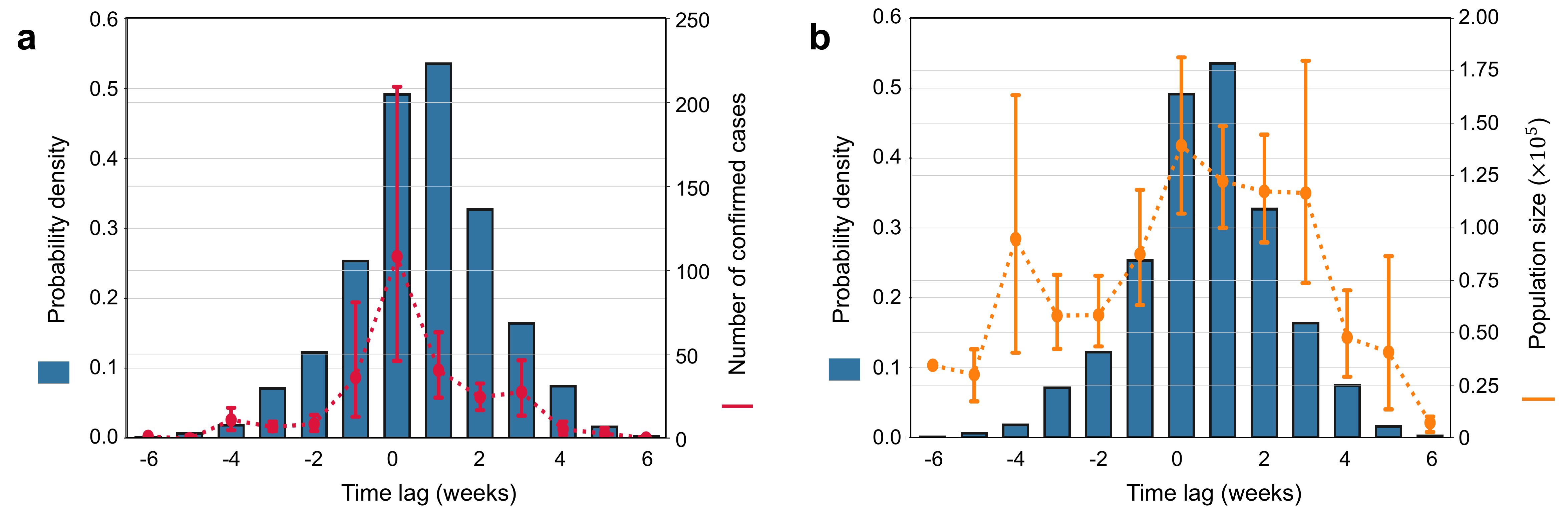}
\caption{Probability density function of time lags between the co-location degree centrality and the second-order growth rate of the cases, (a) with number of confirmed cases on March 31, 2020 (red); and (b) with population size (orange). The error bar represents the two-side $95\%$ confidence interval. Some counties with extreme numbers of confirmed cases are considered outliers and are not included. (Statistics of the time lags: mean: 0.6, median: 1.0, variance: 2.9, $95\%$ confidence interval: [0.53, 0.67]).}
\label{fig_3}
\end{figure}

\subsection*{Time lag between travel reduction and second-order growth rate}
To explore the effect of travel reduction on the growth of weekly new cases in counties, we examined the relation between co-location degree centrality and the second-order growth rate. Based on observations drawn from Fig. \ref{fig_2}, we found that when co-location degree centrality decreased to its lowest level, the second-order growth rate dropped and fluctuated around zero. This result raises an important phenomenon about the synchronicity and time lags between these two variables. 

To quantify the temporal relationship between co-location degree centrality and the second-order growth rate, we identified the week when the degree centrality first reached to its lowest point, and the week when the second-order growth rate first reached to a value below zero for a county. By comparing these two timestamps, we can capture the time lags between reduction of travel and the reduction of weekly new cases for a county. Fig. \ref{fig_3} shows the distribution of the time lags for all US counties along with the number of confirmed cases on 31 March 2020 and the population size of each county. We can discern that the mean of the time lag in which the co-location degree centrality reached the lowest value prior to the first negative second-order growth rate is 0.6 weeks, and medium time lag is 1 week. That means, in most counties, the reduction of co-location degree centrality had a delayed effect on the reduction of new weekly cases. In the counties with the greatest number of cases, however, the reduction of degree centrality had no delays (Fig. \ref{fig_3}a). That is, the week when the degree centrality reached the lowest value is the same week when the second-order growth rate became negative. The result implies that the reduction of population co-location has a synchronic effect on the growth of weekly new cases in the majority of severely infected counties but is more likely to have a one-week lag in other counties. By further associating the population size of the counties with the time lags, we found that, with few exceptions, the time lags can be captured by population size of a county (Fig. \ref{fig_3}b). The majority of counties with large populations tend to have synchronic or one-week lagged association between travel reduction and the growth of weekly new cases. Nevertheless, some counties with high population but a small number of cases could have negative growth of weekly new cases prior to the time when the degree centrality reached to its lowest. This pattern might be explained by proactive local social distancing measures in those counties.

\begin{figure}[ht]
\centering
\includegraphics[width=17cm]{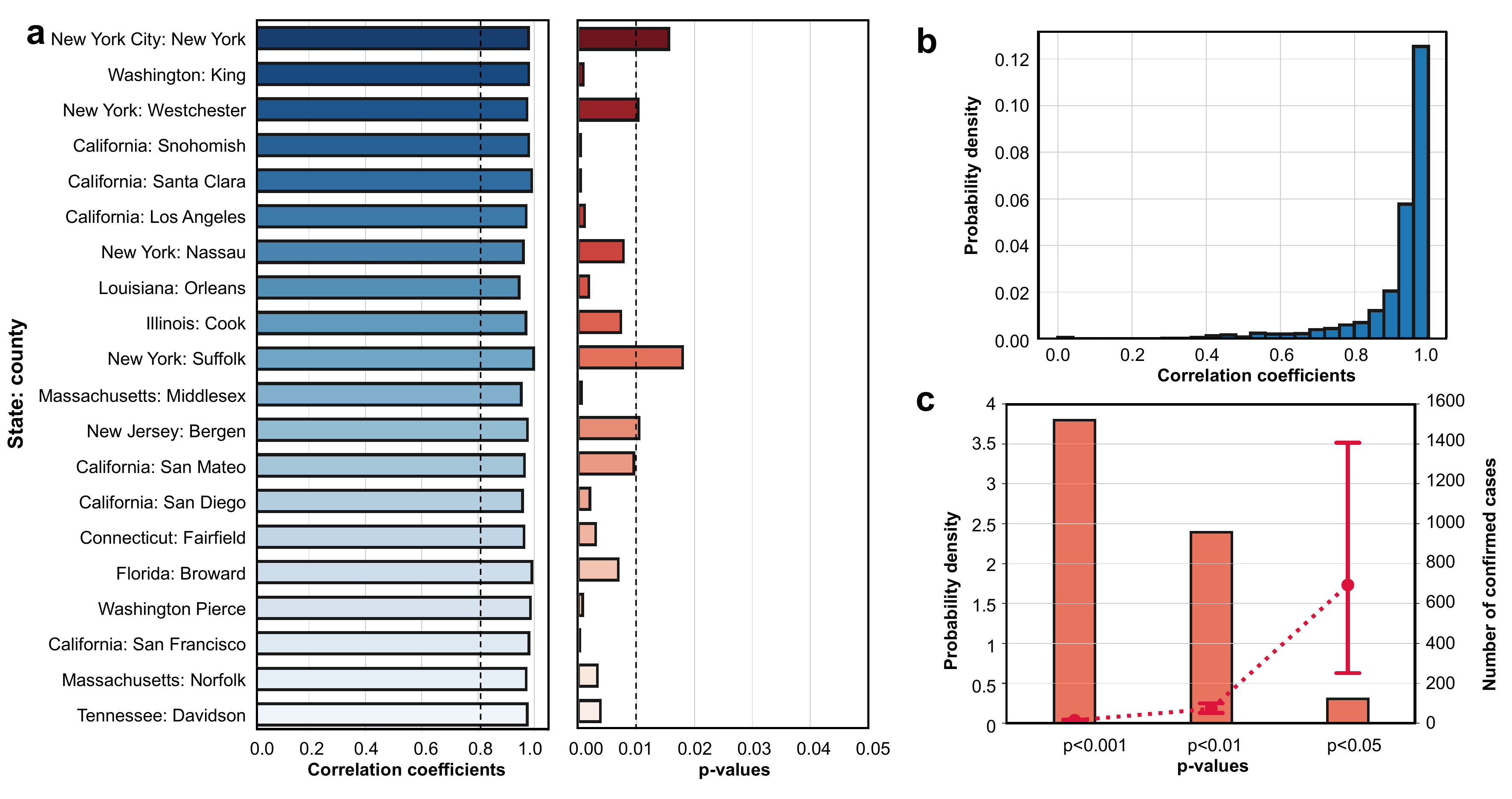}
\caption{(a) Correlation coefficients between co-location degree centrality and estimated basic reproduction number and the p-values for 20 counties with the greatest number of confirmed cases on March 31; (b) distribution of correlation coefficients for all counties; and (c) distribution of p-values (orange) for the correlation analysis for all counties. The data for confirmed cases is based on the CDC reports on March 31, 2020 (red). The error bar represents the two-side $95\%$ confidence interval. Some extreme numbers of confirmed cases are considered as outliers and are not included. }
\label{fig_4}
\end{figure}

\begin{figure}[ht]
\centering
\includegraphics[width=17cm]{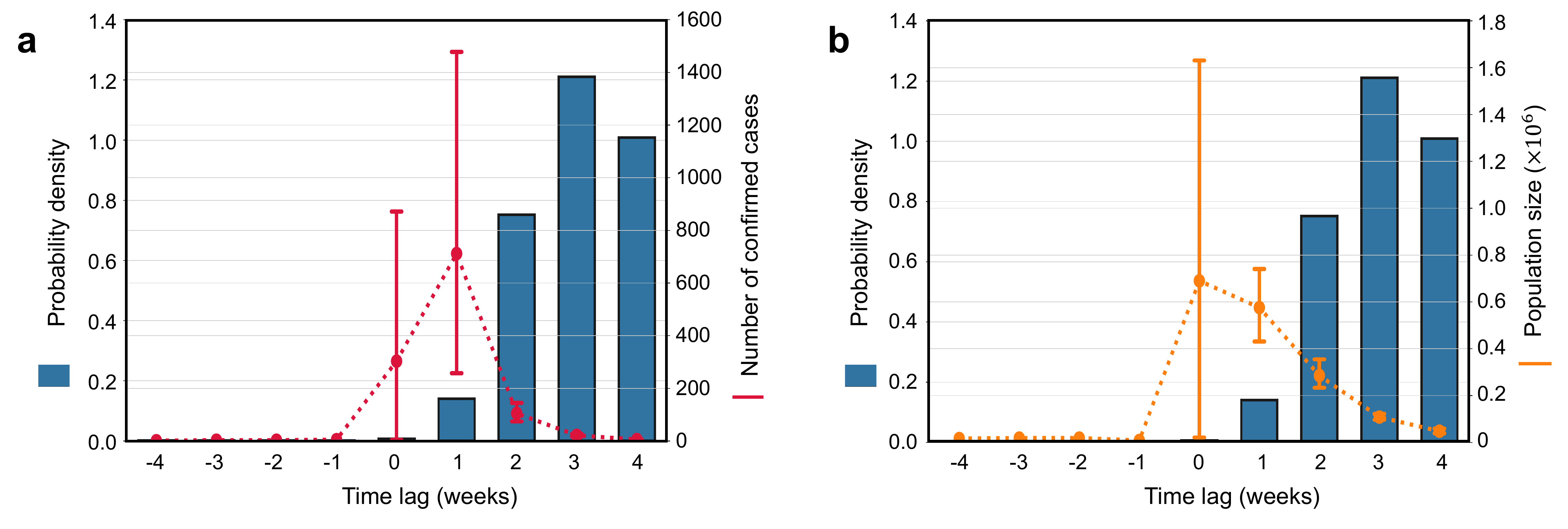}
\caption{Probability density function of time lags between co-location degree centrality and weekly basic reproduction numbers (blue bar plot). (a) with the average number of confirmed cases on March 31, 2020, for the counties in each time-lag group (line plot with error bar); and (b) with the population size of a county. Positive time lag indicates degree centrality decreases prior to the decrease of basic reproduction number (indicating proactive social distancing), while negative time lag indicates the basic reproduction number decreases prior to the decrease of degree centrality (indicates obeying stay-at-home orders). The error bar represents the two-side $95\%$ confidence interval. Some extreme numbers of confirmed are considered as outliers and are not included in the analysis. }
\label{fig_5}
\end{figure}

\subsection*{Time lag between travel reduction and the estimated basic reproduction number}
According to observations drawn from Fig. \ref{fig_2}, reduction in the estimated basic reproduction number, to some extent, followed the reduction of co-location degree centrality, but they are not completely synchronic. Quantifying time lags between these two metrics can provide important evidence for assessing and predicting the transmissibility of the disease in counties with susceptible populations. To this end, we employed the time-lagged cross-correlation method to examine the extent to which the time series of travel reduction (measured by the co-location degree centrality) and the reduction of the basic reproduction number are synchronized and the length of the time, in general, between these two metrics (see Materials and Methods for details). In particular, in this study, we mainly focus on the synchronicity of the descending periods of these two metrics.

Fig. \ref{fig_4}a shows the correlation coefficients and corresponding significant levels for the top 20 counties with the highest number of confirmed cases based on the CDC reports on March 31. We found that the reduction of co-location degree centrality has a significant positive correlation with the estimated basic reproduction number in the counties with the greatest number of cases. The positive correlation means that the decrease in travel is associated with the decrease in the basic reproduction number. Furthermore, the synchronicity analysis enables us to identify the time lags by moving the curve of one metric and to find the highest correlation coefficients. By offsetting the time lags between two variables, we find that the majority of the counties have significant positive correlation between the co-location degree centrality and the basic reproduction number (Fig. \ref{fig_4}b and \ref{fig_4}c). The correlation is more significant especially in the counties with small number of infected population (Fig. \ref{fig_4}c). These results indicate that people reduced their cross-county travel proactively when they became aware of the risks of COVID-19 and when stay-at-home orders were issued. Then, the transmissibility of the pandemic (reflected by the estimated basic reproduction number) reduced accordingly. But, the effect of travel reduction on decreasing the transmissibility might appear with delays.

The length of the time lag is rather important for monitoring and tracking the effect of population co-location and travel reduction on the transmission of COVID-19. To uncover the time lag across different counties, we also examined the distribution of counties in terms of the time lags in Fig. \ref{fig_5}. The results show that, the reduction of co-location degree centrality in counties with the greatest number of infectious cases occurred about one week prior to the reduction of the basic reproduction number in their counties. This could imply that the effects of travel reduction (stay-at-home action) would appear with about a one-week delayed effect on the reduction of the transmissibility of the disease (basic reproduction number). On the other hand, in the counties with a smaller number of infected populations, the occurrence of travel reduction follows an extreme case. People in some counties with low percentages of infected populations reduced their travel proactively, but the effect of travel reduction might appear with a delay of more than two weeks. This lag could be attributed to some other factors, such as community spread and local contact activities. By further associating the population size of a county with the time lags, we find that counties with the greatest population size tend to have a synchronic relationship between co-location degree centrality and the basic reproduction number, or the co-location degree centrality might decrease one week prior to the basic reproduction number. Hence, the population size of a county could be related to the time lags and become an indicator of the effect of travel reduction on diminishing the basic reproduction number. The results confirm that travel reduction in counties with a high number of cases and a large population size would be more effective in reducing the transmissibility of the disease compared to other counties.

\begin{figure}[!htb]
\centering
\includegraphics[width=17cm]{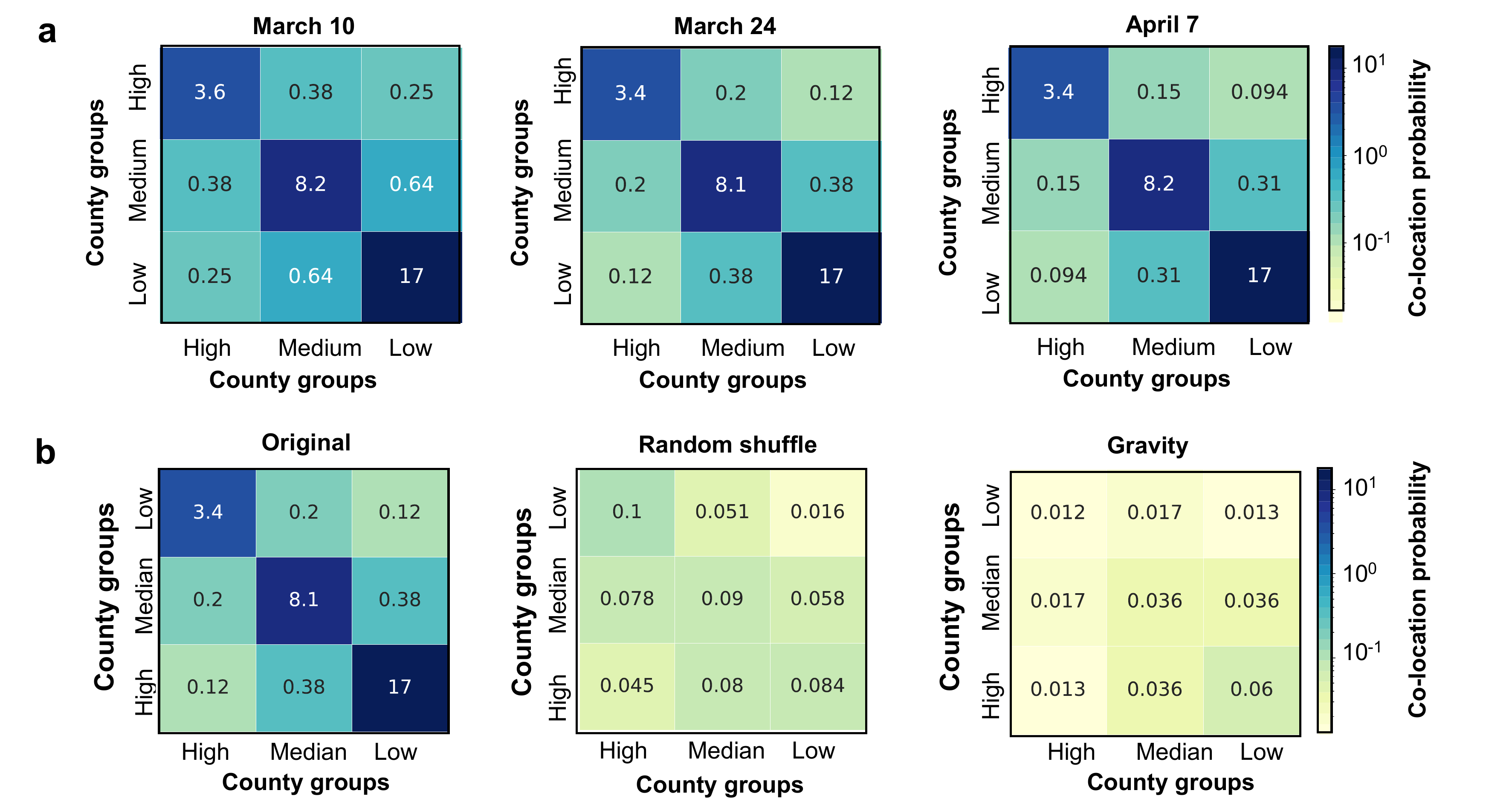}
\caption{(a) Co-location probability across different county groups for three different time periods. (b) The results generated were based on original data (24 March 2020), random shuffle, and gravity model with input of population size for testing the significance of the segregation of co-location patterns based on travel across different county groups. The value in each cell is the sum of the co-location probabilities of the same type of edges. High, medium, and low groups are the groups of counties which are ranked by the number of confirmed cases on 31 March 2020 (the ranks based on the data in different weeks were performed as a sensitivity test provided in Supplementary Information). The sizes of the groups are equal (about 1,000 counties per group). The color bar is consistent for the heat maps in the same subplot.}
\label{fig_6}
\end{figure}

\begin{figure}[!htb]
\centering
\includegraphics[width=17cm]{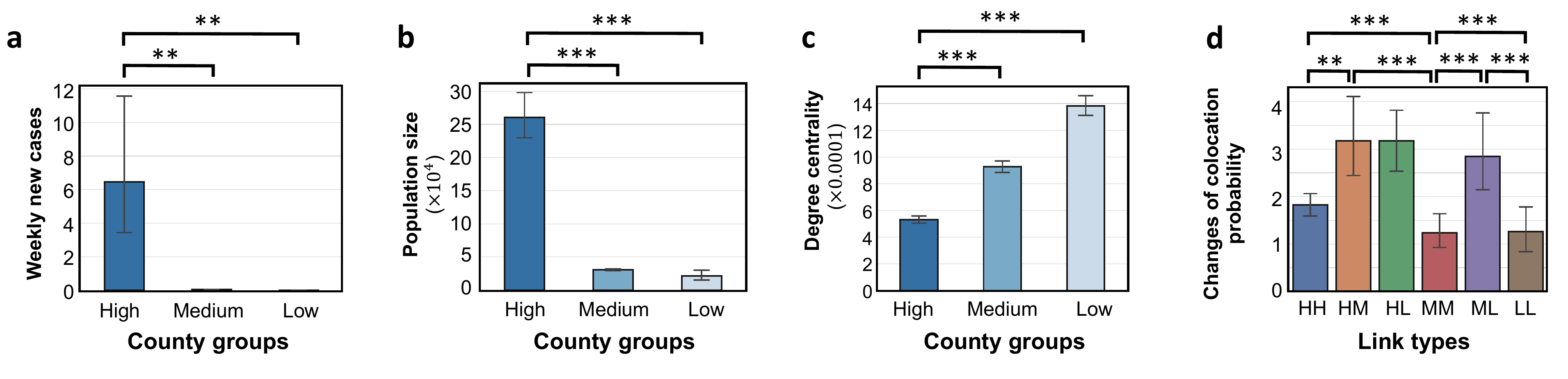}
\caption{Two sample t-tests for the metric values across three different county groups. The metrics are: (a) weekly new cases (first-order growth rate) over the week from March 17 to March 24; (b) population size; (c) co-location degree centrality over the week March 31; and (d) the percentage reduction of the co-location probability for different types of links over the week from March 10 to March 17. HH means the edges connecting two counties both from the high-risk group; HM means the edges connecting one county from high risk group and another county from medium risk group, and so on so forth. (the ranks based on the data in different weeks were performed as a sensitivity test provided in Supplementary Information). Note: $*** p<0.001$; $** p<0.01$; $* p<0.05$.}
\label{fig_7}
\end{figure}

\subsection*{Travel reduction and segregation across county groups}
Due to the dependence of virus transmission on human travel, the spatial patterns of cross-county travel activities should be explored. To distinguish the counties with varying infection situations, we grouped the counties into three categories: high, medium, and low, based on the rank of the counties by the number of confirmed cases on March 31. Then, we summed up all the edge weights connecting the counties within or across groups to evaluate the co-location probability of each county with other counties within the same group. We found that within-group travels led to the highest co-location probability among people from counties in the same group (Fig. \ref{fig_6}a). In order words, people have a greater co-location probability with people from counties with a similar number of cases. The result indicates a segregation among the counties from different groups at different weeks during the initial onset of the outbreaks. This raises important questions: whether travel segregation across county groups with different levels of pandemic situations is an inherent segregation pattern, and whether the pattern can be captured by geographical segregation and population size of two paired counties. 

To test the significance of the co-location segregation, for comparison, we showed the simulated results from two artificial co-location networks generated by (i) randomly shuffling the counties’ pandemic situations in a null model; (ii) simulating the co-location values with a gravity-based model in which the attractive attribute of a county is the population size. (See Materials and Methods for more details.) Considering March 17 travel patterns as an example (Fig. \ref{fig_6}b), we can observe that the segregation (observed from the Facebook co-location data) is greater than the one produced by the null model (random shuffle model) showing that the pattern is significant and not an artifact result. Second, the segregation is stronger than that predicted based solely on geographic attributes (using the gravity model), indicating that the observed segregation cannot be simply attributed to the size of population in the counties. 

Due to the presence of significant co-location segregation due to travel activities, it is also necessary to examine some attributes which vary across the three county groups and could be related to the transmission of COVID-19. In this study, we focus mainly on three attributes: growth rate in number of weekly new cases (Fig. \ref{fig_7}a), population size (Fig. \ref{fig_7}b), and co-location degree centrality (Fig. \ref{fig_7}c). The growth rate of the infected population in a county is the average number of new confirmed cases in the week March 17 to 24. We find that the group of counties with the highest number of confirmed cases tend to have high growth rate (Fig. \ref{fig_7}a). This is quite intuitive since the counties with more infected people would increase the chances of community transmission in a county. Additionally, the group of counties with the highest number of confirmed cases also has large populations, compared to counties with medium and low numbers of infections (Fig. \ref{fig_7}b). Due to their large populations, counties with the highest number of confirmed cases have low co-location degree centrality, indicating that persons from these counties have less co-location probability than the persons from counties with smaller populations (Fig. \ref{fig_7}c).  This result, however, accounts only for cross-county travel patterns. Disease transmission should include both community spread and travel-related cross-region transmission. This study only focuses on transmission risks among counties to reveal a cross-country transmission risk patterns based on human travel activities. 

By further comparing the travel patterns across the weeks in March and April, we found that within-group co-location was always dominant and remained stable over time, while cross-group co-location was reduced about $50\%$ between March 10 to March 24, and continued to decrease by $18\%$ in the following two weeks (Fig. \ref{fig_6}a). We also conducted a significance test for within-group and cross-group travel reduction based on the types of the edges. Fig. \ref{fig_7}d shows that, from the same groups, the reduction of co-location probability on the edges connecting the counties is significantly less than the edges connecting the counties from different groups, meaning that social distancing was not as extensively performed among people from the counties within the same group. For example, the co-location probability among people from counties with the greatest number of cases did not change extensively. Nevertheless, the segregation of co-location probability for counties within the same group contributes to isolating the counties by groups so that the disease would be less likely to be transmitted from one group to another. For example, counties with low numbers of cases have a lower transmission risk due to travel-related spread from highly inflected counties.

\section*{Discussion}
This study provides quantitative empirical evidence regarding the relationship between population co-location and travel reduction and the spatiotemporal transmission risk of COVID-19 in the United States. using the Facebook co-location maps. The results regarding both temporal and spatial patterns of travel reduction could provide worthy implications for epidemic models and policies to control the transmission risk of COVID-19 and other future pandemics. 

We analyzed the synchronicity between travel reduction and the growth rate of weekly new cases (second-order growth rate) for each county, and found that travel reduction has a synchronic effect on the reduction of weekly new cases in the counties with greatest number of cases, while showing an average one-week lag in the majority of other counties. This finding indicates that reducing the population co-location and cross-county travel has a positive effect on reducing the growth rate of new weekly cases. This effect is more prominent in counties with greater population size. In addition, in examining the synchronicity between travel reduction and the estimated basic reproduction number, we also found that reduction in the basic reproduction number tends to be synchronized to travel reduction with a one-week lag. The synchronicity is more evident in counties with the largest population sizes. Considering these two noteworthy empirical findings, we can confirm the importance of travel reductions, specifically in counties with large population size, in containing the growth of an epidemic. The effect of travel reduction may not show immediately but rather gradually reveal itself with a one-week lag. Of particular note, local governments can project the number of weekly new cases and the basic reproduction number from at least one week after the implementation or lifting of social distancing orders to assess the effectiveness and necessity of the order. If co-location degree centrality of a county grows after lifting social distancing orders, the growth rate of number of weekly new cases may follow. 

We also investigated travel patterns among county groups and found segregation between the counties from different groups (categories based on number of confirmed cases). In particular, within-group travel was far more prominent than cross-group travel and did not change significantly after stay-at-home orders were issued. Such a segregated travel pattern might have been beneficial to control the spread of COVID-19 across different groups but has potential to exacerbate the transmission in highly infected counties. In addition, we also found that these highly infected (large numbers of infected population) counties tend to have large population sizes and growth rates of the infected populations. These attributes enable these counties to be hotspots which have higher risks of contamination\cite{Oliver2020}. These hotspots with a high level of interaction and a higher concentration of population become the epicenters of the pandemic spread. We also found that the counties with medium and low levels of infected populations have even higher co-location probabilities than the counties with high levels of infected populations. The empirical evidence regarding the existence of segregation in cross-county co-location patterns could not only inform the epidemiologic modeling for the transmission of COVID-19, but also have practical implications. Specifically, the presence of segregation across different county groups can enable accurate projection of the trajectory of new infection cases for purposes of crafting policies for enforcing and relaxing travel restrictions. By comparing the travel reduction over time for different types of edges, we found that within-group travel remains stable but cross-group travel decreased by more than $60\%$. The heterogeneity of travel reduction reveals that social distancing was not well practiced for within-group travels, which potentially could contribute to travel-related spread of the virus among populations of counties with a high number of cases. Conversely, social distancing led to a reduction in cross-group travel, which contributes to a reduction in cross-group spread of COVID-19. This result reveals that social distancing orders do not have a homogenous effect on travel reduction across all counties. This phenomenon is overlooked in most mathematical models and policy-making processes.

Although the findings in this study provide useful theoretical and practical implications for disease control, a few limitations in this study should also be noted. First, this study only analyzed the situation from March until early May 2020. As the pandemic continues, the synchronicity analysis among the metrics in this study could be further tested and the changes under various situations (such as business reopening and lifting of social distancing orders) can be examined. In addition, this study relies primarily on Facebook co-location maps in which the mobility data is collected and generated based on the activities of the Facebook users and their geographical location services. To examine the generality of this results, future studies can also employ other data (such as mobile data) to validate the patterns identified in this study. Finally, the present study focuses on US counties during the period preceding and following the outbreaks of COVID-19. It will be important to explore the co-location patterns and the relationship with the cross-region epidemic spread in other countries.

\section*{Materials and Methods}
\subsection*{Datasets and preprocessing}
We collected the Facebook co-location maps spanning 10 weeks from the span 3 March to 5 May 2020. Each weekly co-location map has about 3,000 nodes that represent the counties across the United States. These maps estimate the probability of a randomly chosen person from two different counties being located in the same place during a randomly chosen minute for a given week\cite{FacebookDataforGood2020a}. Same place here roughly corresponds to a 0.6 km by 0.6 km square depending on the latitude in the Microsoft Bing tile system\cite{Qu2011a}. The co-location maps were generated on a weekly basis in which the computation of co-location probabilities for each pair of counties are independent across the weeks, except the bounding boxes for identifying the same places. Facebook generated the co-location maps from the mobility data of 27 million Facebook mobile app users with location history turned on\cite{Maas2019}. Mobility data are location updates from a user’s cell phone in the form of latitude and longitude at a given time. Facebook assigned the mobility data to counties and computed the co-location probability for each pair of counties. 

The infected population data we used in this study were gathered from the COVID-19 Data Repository by the Center for Systems Science and Engineering (CSSE) at Johns Hopkins University\cite{JohnHopkinsUniversity2020}. The data were reported daily by the Centers for Disease Control and Prevention (CDC) for all counties in the United States. The CDC data set includes the cumulative number of confirmed cases and deaths, and the FIPS code (Federal Information Processing Standards code) of each county from late January 2020 until early May 2020. This study looked only into the growth of confirmed cases and their spatial distribution at the county level. Hence, we adopted the data for each week corresponding to the weeks defined by Facebook and calculated the weekly new confirmed cases for each county. By doing so, we can obtain the time series of confirmed cases for weeks during the period of interest and for all counties in the United States. Furthermore, the growth rate of weekly new confirmed cases can be calculated by: 
\begin{equation}
    \centering
    v_{i,j}=c_{i,j}-c_{i-1,j}
\end{equation}
where $v_{i,j}$ is the growth rate of weekly new cases (second-order growth rate) for week $i$, and county $j$, $c_{i,j}$ is the weekly new cases in week $i$ and county $j$. 

In addition to the co-location maps and CDC data, we also included a feature of population size of each county in our analysis. The population data are the 5-year estimates for the 3,142 counties in the United States based on 2014–2018 American Community Survey (ACS), released recently by the US Census Bureau\cite{UnitedStatesCensusBureau2019}. We associated the CDC data, co-location probability, and the population data together with the FIPS code of each county. 

\subsection*{Co-location weighted spatial network}
The spatial network studied here is converted from the weekly co-location maps. The network has about 3,000 nodes and 1 million undirected edges in the week ending 24 March 2020 during the outbreak of COVID-19. The nodes represent the US counties with mobility data recorded by Facebook. The edges are weighted by the standardized co-location probability for each pair of counties. With a proper assignment of weights to the edges, we can characterize cross-county transmission risks due to travel patterns of the people in a county by computing its degree centrality, which could inform the co-location patterns of the people in a county with people from other counties. The degree centrality\cite{barabasi2016network} for each node in the weighted network can be obtained by:
\begin{equation}
    \centering
    k_{i}=\sum_{j=1}^{n}w_{ij}
\end{equation}
where $k_i$ is the degree centrality of node $i$, $w_{ij}$ is the weight of the edge connecting node $i$ and node $j$, and $N$ is the number of nodes in the network $G$.

\subsection*{Estimation of basic reproduction number}
Basic reproduction number ($R_0$) is defined as the expected number of secondary cases produced by a single infection in a susceptible population\cite{Dietz1993}. Due to a lack of official estimation and report about this basic reproduction number, we estimated this metric for each county based on the growth of confirmed cases using a simple epidemic model. First, we assume that an individual infects an average of $R_0$ new individuals after exactly a time $tau$ (the serial interval) has passed. Then starting with $i(0)$ individuals, the number of infected individuals will be $i(t)=i(0)R_0^{t/{\tau}}$\cite{McCluskey2010}. Hence, by taking logarithm of both sides of the equation, $R_0$ can be estimated by
\begin{equation}
    \centering
    \ln{R_0}=\frac{\ln{i(t)}-\ln{i(0)}}{t/\tau}
\end{equation}
Simplifying the equation with a substitution, we can obtain that 
\begin{equation}
    \centering
    R_0=e^{K\tau}
\end{equation}
where $K=({\ln{i(t)}}-{\ln{i(0)}})/t$. A simple epidemic contagion model tends to assume that growth of an epidemic in the early stages is exponential in a short period\cite{He2020}. Hence, we adopted CDC data from the date of interest to ten days prior to this day and calculated the values of $K$ for each county per day. Based on the existing studies and models on COVID-19\cite{Zhang2020,Zhang2020a}, we set ${\tau}$ to be $5.1$ days. Then, we can estimate $R_0$ for each county per day using equation (4).

\subsection*{Time-lagged cross correlation}
The time-lagged cross-correlation enables measurement of the similarity of two time series data sets as a function of displacement of one relative to another. This method has been widely used in signal processing and pattern recognition\cite{Hernandez-Varas2015}. It leverages the time-resolved information contained in the synchronous dynamics of two variables. In this study, we used this method to analyze the synchronicity and time lag between travel reduction and growth rate of infected population. Although we can observe similar trends between travel reduction and growth rate of infections, these two time series variables do not change at the same time. To offset the time lag, we incrementally shift one variable (travel reduction) and repeatedly calculate the correlation between two variables. The formula we adopted to obtain the cross-correlation is shown as below\cite{Gubner2006}:
\begin{equation}
    \centering
    \rho_{XY}(\tau)=\frac{E[(X_t-\mu_X) \overline{(Y_{t+\tau}-\mu_Y)}]}{\sigma_X \sigma_Y}
\end{equation}
where $\rho_{XY}$ is cross-correlation for time series $X$ and $Y$, $\tau$ is the displacement which can be considered as the offset of two variables, $X_t$ and $Y_t$ are two time series variables, $\mu_X$ and $\sigma_X$ are the mean and standard deviation of the process $X_t$ at time $t$, $\mu_Y$ and $\sigma_Y$ are the mean and standard deviation of the process $Y_t$ at time $t$, $E[]$ is the expected value operator, and  $\overline{(Y_{t+\tau}-\mu_Y)}$ denotes the complex conjugate of $(Y_t-\mu_Y)$. Here, $\rho_{XY}$ is well defined, and its value must be in the range of $[-1,1]$, with $1$ representing perfect positive correlation, and $-1$ representing perfect negative correlation. Using this function, we can identify the offset that can maximize the cross-correlation of two time series variables. 

\subsection*{Artificial co-location maps}
To test the significance of the travel segregation patterns, we created two artificial co-location networks\cite{Dong2019a}. The first artificial network is a null model, in which the number of confirmed cases is randomized to leave only the co-location probability for each pair of counties. To do that, we first extract a list of counties included in this study and their corresponding number of confirmed cases on 31 March 2020. Then, we randomly shuffle the number of confirmed cases and associated the values with each county. Using this data, we finally plot the heatmap shown in Fig. \ref{fig_6}b.

To illustrate the difference between the empirical evidence and the one caused by population distribution, the second artificial co-location network is generated by a gravity-based model which incorporate the population size of a county and the distance between two counties\cite{Simini2012,Bassolas2019}. We define the co-location metric value between county i and j as the dependent variable $T_{ij}$. Then, using the population size at the attractive attribute of two counties, we can predict the $T_{ij}$ with the following formula:
\begin{equation}
    \centering
    T_{ij}=k\frac{V^{\mu}W^{\alpha}}{d^{\beta}}
\end{equation}
where $k$ is a scaling factor that can be estimated from the empirical data to ensure the total observed and predicted metric values are consistent, $V$ is the value of attractive attribute of the origin, $W$ is the value of the attractive attribute of the destination, $\mu$ is the parameter which controls the effect of the origin attribute, $\alpha$ is the parameter which controls the effect of the destination attribute, and $\beta$ is a parameter representing the effect of distance between the origin and destination. 

To fit this model, we converted the equation (6) to equation (7) by taking the logarithm on both sides of the equation. 
\begin{equation}
    \centering
    \ln{T_{ij}}=\ln{k}+\mu\ln{V}+\alpha\ln{W}-\beta\ln{d}
\end{equation}
Then, we adopted the linear regression model to estimate the parameters: $k$, $\mu$, $\alpha$ and $\beta$. Both on the fitted parameters, we can compute the metric value $\tilde{T_{ij}}$ for any pair of counties with their population sizes and distance. Based on predicted metric value, we can plot the heatmap as shown in Fig. \ref{fig_6}b to examine the segregation in human travel activities.

\bibliography{Colocation_paper}

\section*{Acknowledgements}
This material is based in part upon work supported by the National Science Foundation under Grant Number SES-2026814 (RAPID), the National Academies’ Gulf Research Program Early-Career Research Fellowship and the Amazon Web Services (AWS) Machine Learning Award. The authors also would like to acknowledge the data support from Facebook Data for Good. Any opinions, findings, and conclusions or recommendations expressed in this material are those of the authors and do not necessarily reflect the views of the National Science Foundation, Amazon Web Services and Facebook.

\section*{Author contributions statement}
Research design and conceptualization: C.F., S.L., A.M.; Data collection, processing, analysis, and visualization: C.F., S.L., Y.Y., B.O., Q.L.; Writing: C.F., A.M.; Reviewing and revising: all authors.

\section*{Competing interests}
Authors have no competing interests.

\section*{Data and materials availability}
The data that support the findings of this study are available from Facebook, but restrictions apply to the availability of these data, which were used under license for the current study, and so are not publicly available.

\end{document}